\documentclass[conference]{IEEEtran}
\IEEEoverridecommandlockouts
\usepackage{cite}
\usepackage{amsmath,amssymb,amsfonts}
\usepackage{algorithmic}
\usepackage{graphicx}
\usepackage{textcomp}
\usepackage{xcolor}
\usepackage{multicol}
\usepackage{multirow}
\usepackage{subfigure}
\usepackage{booktabs}
\def\BibTeX{{\rm B\kern-.05em{\sc i\kern-.025em b}\kern-.08em
    T\kern-.1667em\lower.7ex\hbox{E}\kern-.125emX}}
\begin{document}

\title{Multi-source Knowledge Enhanced Graph Attention\\ Networks for Multimodal Fact Verification}

\author{\IEEEauthorblockN{Han Cao\IEEEauthorrefmark{1}\IEEEauthorrefmark{2},
Lingwei Wei\IEEEauthorrefmark{1},
Wei Zhou\IEEEauthorrefmark{1}, and
Songlin Hu\IEEEauthorrefmark{1}\IEEEauthorrefmark{2}
}
\IEEEauthorblockA{\IEEEauthorrefmark{1}Institute of Information Engineering, Chinese Academy of Sciences, Beijing, China }
\IEEEauthorblockA{\IEEEauthorrefmark{2}School of Cyber Security, University of Chinese Academy of Sciences, Beijing, China}
\thanks{Corresponding author: Lingwei Wei (email: weilingwei@iie.ac.cn).}}

\maketitle

\begin{abstract}
Multimodal fact verification is an under-explored and emerging field that has gained increasing attention in recent years. The goal is to assess the veracity of claims that involve multiple modalities by analyzing the retrieved evidence. The main challenge in this area is to effectively fuse features from different modalities to learn meaningful multimodal representations. To this end, we propose a novel model named Multi-Source Knowledge-enhanced Graph Attention Network (MultiKE-GAT). MultiKE-GAT introduces external multimodal knowledge from different sources and constructs a heterogeneous graph to capture complex cross-modal and cross-source interactions. We exploit a Knowledge-aware Graph Fusion (KGF) module to learn knowledge-enhanced representations for each claim and evidence and eliminate inconsistencies and noises introduced by redundant entities. Experiments on two public benchmark datasets demonstrate that our model outperforms other comparison methods, showing the effectiveness and superiority of the proposed model.
\end{abstract}


\section{Introduction}
Fact verification, aiming to assess the truthfulness of claims by the retrieved evidence, has attracted a great amount of attention in research fields \cite{GuoSV22}.
With the increasing availability of multimedia data, \textit{Multimodal Fact Verification} has emerged as a new research direction, necessitating a comprehensive understanding and integration of different modalities to make accurate predictions \cite{Roy21}.

Recent approaches for multimodal fact verification have been proposed. They utilize textual and visual content and leverage different neural networks to learn modality-specific features from claims and evidence \cite{Roy21, Singhal22, Wu21, Qian21}. To comprehensively incorporate multimodal features, diverse fusion methods are designed to integrate multimodal representations for verification. Some approaches \cite{Mishra22, Gao22, Dhankar22} utilize concatenation operations to fuse features from different modalities. To learn the interaction between modalities, some works \cite{Wu21, Wang22, Singhal20} leverage cross-attention mechanisms to integrate multimodal embeddings. 
\begin{figure}[t]
  \begin{center}
  \includegraphics[width=0.9\linewidth]{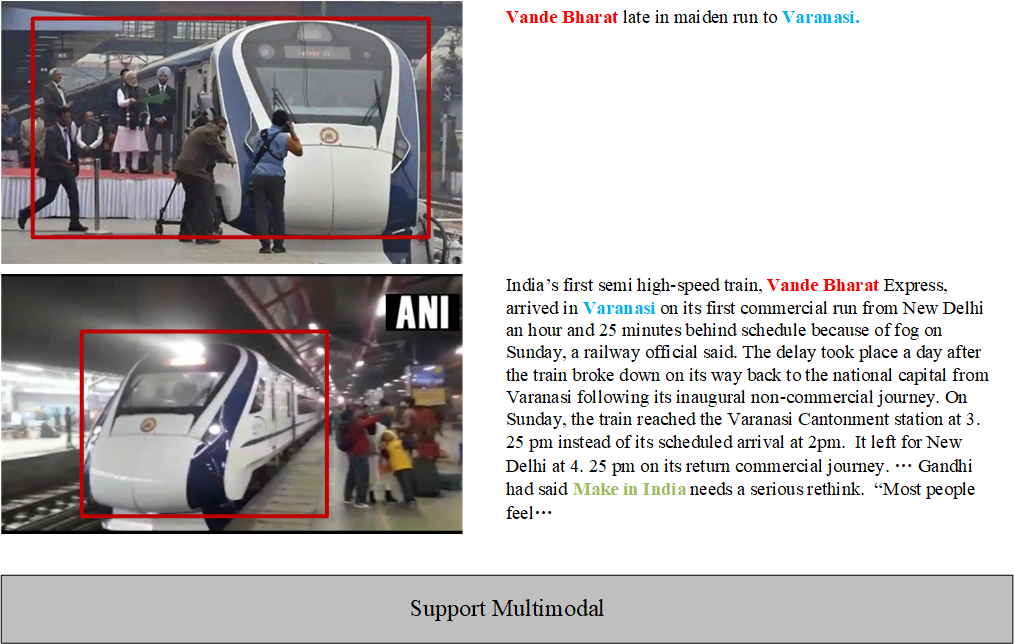}
  \end{center}
  \caption{Examples of the effectiveness of entities in multimodal fact verification. \textit{Support} means the entailment of claim and evidence. \textit{Multimodal} means similar multimodal contents.}
  \label{f:1}
\end{figure}

However, recent studies rely on coarse-grained rationales, such as sentence-level information \cite{Roy21, Qian21}, and fuse multimodal information by concatenation \cite{Gao22, Dhankar22} or attention-based modules \cite{Wu21, Wang22, Singhal20}, which overlooks the important details contained in fine-grained entities and neglect complex interactions between different modalities. Besides, fine-grained knowledge from different sources contains plenty of semantic information and can complement each other to enhance the data from different perspectives.
As shown in Fig.~\ref{f:1}, these two images of claim and evidence are matched for their main characters marked with red rectangles are the same, and so are the textual contents marked with red and blue. It has been demonstrated that a more detailed understanding of the texts and images involved can be obtained by considering specific entities and objects in text and images, which leads to more accurate judgments regarding the veracity of multimodal claims\cite{QiCao21}. Besides, Large Language Models (LLMs) can also extract key information and evidence as augmented data in the fact verification tasks to improve the performance \cite{tan2023evidencebased}. These motivate us to propose the first research question of this paper: {How can knowledge from different modalities and different sources be effectively integrated to understand better and judge the truthfulness of facts?} 

Besides, the entity \textit{Make in India} marked with green in Fig. \ref{f:1} has no contribution to the fact verification task since neither the claim text nor the image mentions it at all, which elucidates that introducing entities inevitably brings in noise that may hinder the model. This motivates us to propose the second research question: {How can inconsistencies and noise across different modalities be eliminated to ensure the accuracy of the predictions?} 

To tackle the aforementioned issues, we propose a novel Multi-source Knowledge-enhanced Graph Attention Network (MultiKE-GAT), to effectively model fine-grained semantic features for improving performance.
Specifically, MultiKE-GAT introduces two kinds of fine-grained external knowledge, textual and visual entities extracted by the toolkit and key information extracted by LLMs, and takes advantage of graph structure that can inherently capture complex intra- and inter-modal relations to learn better multi-source knowledge-enhanced representations.
Moreover, MultiKE-GAT introduces a novel Knowledge-aware Graph Fusion module, which employs global representations as the guide to emphasize relevant entities and marginalize the insignificant ones, to alleviate the negative impact of noise and inconsistency.
To evaluate the performance of our model, we conduct several experiments on two public multimodal fact 
verification benchmarks \cite{Mishra22, DBLP:conf/sigir/YaoS0CH23}. We also carry out ablation studies to further testify to the effectiveness of each module of our model.

\begin{figure*}[htbp]
  \begin{center}
  \includegraphics[width=0.94\linewidth]{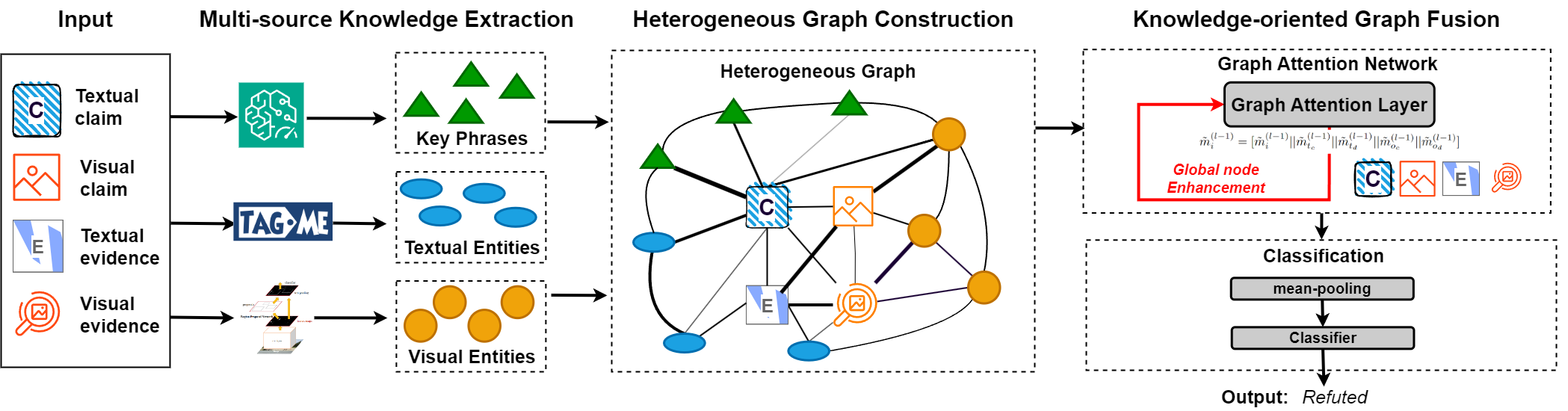}
  \end{center}
  \caption{The overall architecture of MultiKE-GAT. First, multi-source knowledge such as textual entities, visual objects, and keyphrases are extracted from texts and images, forming an undirected heterogeneous graph. The knowledge-oriented graph fusion network fuses diverse fine-grained knowledge with the guidance of global nodes to learn multimodal representations. We use the representation as input to the MLP-Classifier for verification.}
  \label{f:2}
\end{figure*}

Our major contributions are as follows:
    {1)} We propose a novel {Multi-source Knowledge-enhanced Graph Attention Network} to integrate multi-granularity semantics to enhance multimodal fact verification. To our knowledge, this is the first preliminary attempt to investigate multi-source fine-grained knowledge for multimodal fact verification tasks.
    {2)} To eliminate the negative impact of introduced noise, we design a new {Knowledge-oriented Graph Fusion module}. It enables to decrease in the risk of amplifying noises with the guidance of global representation.
    {3)}  To evaluate the performance of the proposed method, we carry out experiments on {FACTIFY} and {Mocheg}. Our model outperforms the comparison methods, which demonstrates the effectiveness and superiority of the proposed model.

\section{Methodology}

\label{sec:Methodology}
Let $\mathcal{C}=\{C_T, C_I, E_T, E_I\}^{|\mathcal{C}|}$ be the corpus of the dataset, where $C_T$ and $C_I$ denote the text and image of the claim, and $E_T$ and $E_I$ denote the text and image of the evidence. Multimodal fact verification aims to find a function $f: \mathcal{C}\rightarrow \mathcal{Y}$ that maps the data to the label set and makes predictions, where $y\in \mathcal{Y}$ denotes the label.

In this section, we propose a novel \textit{Mulitsource Knowledge-enhanced Graph Attention Network} model (MultiKE-GAT) to capture more granular information in entities and objects in textual and visual content for fact verification. Fig. \ref{f:2} illustrates the overall architecture of MultiKE-GAT.

\subsection{Multi-source Knowledge Extraction}
\label{sec: Ent-extract}

We first extract multi-source external knowledge like textual and visual entities and key information utilizing an entity extraction toolkit and LLMs.
For a given claim-evidence pair $C=\{C_T, C_I, E_T, E_I\}$, we utilize a toolkit TAGME \footnote{https://sobigdata.d4science.org/group/tagme/} to extract textual entities $t_i^C$ and $t_j^E$, and leverage Faster R-CNN \cite{Ren15} \footnote{https://github.com/pytorch/vision/tree/main/torchvision/models/detection} to detect visual entities $o_m^C$ and $o_n^E$. To maintain fine-grained knowledge as much as possible and filter noisy entities, we set thresholds to control the scale of introduced knowledge. Besides, we utilize LLM to help extract key information in the claim text and evidence text $k_p^C$ and $k_q^E$. We then obtain the multi-source knowledge sets and remove any repeated elements as duplication can negatively impact the performance, and obtain textual entity set $T=\{t_1, t_2, \cdots, t_{|T|}\}$, visual entity set $O=\{o_1, o_2, \cdots, o_{|O|}\}$, and key information set $K=\{k_1, k_2, \cdots, k_{|K|}\}$, where $|T|$, $|O|$, and $|K|$ denotes the length of these sets, respectively.

\subsection{Heterogeneous graph construction}
\label{sec: HGC}

Based on the above entities, we construct a heterogeneous graph to promote a better combination of textual and visual content. For each sample $C$, we define an undirected graph $G=\{V, E\}$, where $V$ and $E$ refer to node and edge sets.

\textbf{Node.} There are three types of nodes in the heterogeneous graph: global, textual entity, and visual entity. Global nodes consist of $t_{c}$, $t_{e}$, $o_{c}$, and $o_{e}$, which are global textual and visual representations of claim and evidence. The textual entity, visual entity, and key information nodes are obtained from set $T$, set $O$, and set $K$ respectively. We leverage a pre-trained language model to obtain the textual features $\tilde{t}_l$ and $\tilde{k}_m$ and a pre-trained visual model to extract visual features $\tilde{o}_n$. Thus, we merge the two vector sets to obtain the node set $V = \{\tilde{t}_{c},\tilde{t}_{e},\cdots,\tilde{t}_{|T|}, \tilde{k}_1, \cdots, \tilde{k}_{|K|},\tilde{o}_{c},\tilde{o}_{e},\cdots, \tilde{o}_{|O|}\}$, and $|V|=|T|+|K|+|O|+4$. 

\textbf{Edge.} There are two types of edges in the graph: homogeneous edge, connecting two homogeneous nodes, and heterogeneous edge, linking between two heterogeneous nodes. To fully leverage the fine-grained and coarse-grained information, we construct a fully connected heterogeneous graph by setting $E=\{e_{ij}\}$, where $0 \leq i,j \leq |V|$. Taking the relation between nodes into account, we set the \textit{cosine similarity} of each node pair as the edge weight.

\subsection{Knowledge-oriented graph-based fusion}
\label{sec: KGF}
Knowledge-oriented Graph-based Fusion (KGF) module aims to effectively integrate external knowledge and capture complex interactions between entity nodes in the heterogeneous graph. 
We first leverage a shared space mapping module to project the node embeddings into a common latent semantic space to eliminate the {heterogeneous gap} \cite{PengQ19}. Specifically, given an entity graph $G_i$, we project the node features by:
\begin{eqnarray}\label{eq-4}
    \tilde{m}_{t}^{(0)} &=& \sigma(FFN(\tilde{t}_l)), t_l \in \{\tilde{t}_{c},\tilde{t}_{e}, \tilde{t}_{1},\cdots,\tilde{t}_{|T|}\}, \\
    \tilde{m}_{k}^{(0)} &=& \sigma(FFN(\tilde{k}_m)), k_m \in \{\tilde{k}_{1},\tilde{k}_{2},\cdots,\tilde{k}_{|K|}\},\\
    \tilde{m}_{o}^{(0)} &=& \sigma(FFN(\tilde{o}_n)), o_n \in \{\tilde{o}_{c},\tilde{o}_{e}, \tilde{o}_{1},\cdots, \tilde{o}_{|O|}\},
\end{eqnarray}
where $FFN$ is the fully connected layer, $\sigma$ is the activation function, and $\tilde{m}^{(0)}=\{\tilde{m}_{1}^{(0)}, \tilde{m}_{2}^{(0)}, \cdots,\tilde{m}_{|V|}^{(0)}\}$ is the input node representation of graph attention layer.

During the graph fusion process, each node embeddings $\tilde{m}_i^{(l)}$ are iteratively updated to capture complex interactions between different types of nodes and edges in the heterogeneous graph. The equation for the $l$-th layer is given by:
\begin{eqnarray}\label{eq-6}
    \tilde{m}_i^{(l)} = \gamma_{i,i}\Theta \tilde{m}_i^{(l-1)} + \sum_{j\in \mathcal{N}(i)}\gamma_{i,j}\Theta \tilde{m}_j^{(l-1)},
\end{eqnarray}
where $\Theta$ denotes the transformation parameters and $\gamma_{i,j}$ is the attention score between node $i$ and its neighbor node $j$:
\begin{eqnarray}\label{eq-7}
    \gamma_{i,j} = \frac{\text{exp}(a^T\text{LeakyReLU}(\Theta[\tilde{m}_i||\tilde{m}_j]))}{\sum_{k\in \mathcal{N}(i)\cup \{i\}}\text{exp}(a^T\text{LeakyReLU}(\Theta[\tilde{m}_i||\tilde{m}_k]))},
\end{eqnarray}
where $||$ is the concatenation operation, LeakyReLU stands for Leaky Rectified Linear Unit.

To alleviate the impact of irrelevant information and noise, we highlight the effect of each global node by concatenating them to each node feature before each convolution layer. 
Thus, before each update process, the node representations are modified by:
\begin{eqnarray}\label{eq-9}
    \tilde{m}_i^{(l-1)} = [\tilde{m}_i^{(l-1)}||\tilde{m}_{t_c}^{(l-1)}||\tilde{m}_{t_d}^{(l-1)}||\tilde{m}_{o_c}^{(l-1)}||\tilde{m}_{o_d}^{(l-1)}],
\end{eqnarray}

Through the above operations, we can automatically obtain fine-grained multimodal representations with less impact of noise and inconsistency.

\subsection{Classifier}
\label{sec: Class}

To predict the label of the given claim-evidence pair, we first aggregate the node features of each graph $G$ using global mean pooling to obtain aggregated representation $r$ as input to the classifier. The prediction process is carried out as follows:
\begin{eqnarray}\label{eq-10}
    r &=& meanpool(\tilde{m}_i^{(L)}),\\
    \hat{y} &=& softmax(W^1\sigma(W^0r)),
\end{eqnarray}
where $L$ denotes the last layer of graph fusion module, $W^0$ and $W^1$ are trainable parameters, and $\hat{y}$ is the predicted label.
We train our model by minimizing cross-entropy loss to learn the prediction of the categories.

\section{Experiment}
\label{sec:exp}

\subsection{Experimental setups}
\textbf{Dataset.}
We conduct experiments on two public benchmark datasets, i.e., FACTIFY \cite{Mishra22} and MOCHEG \cite{DBLP:conf/sigir/YaoS0CH23}.
\textbf{FACTIFY} contains 42,500 multimodal claim-evidence pairs with five categories, i.e., \textit{Support Multimodal, Support Text, Insufficient Multimodal, Insufficient Text}, and \textit{Refute}. We randomly redivide the original training set by 80\% and 20\% as the training and validation set respectively. The original validation set is utilized as the test set. 
\textbf{MOCHEG} contains 21,184 claims and relevant multimodal evidence. It contains three categories, including \textit{Supported, Refuted}, and \textit{Not Enough Info} {({\textit{NEI})}. We utilize the original partition as the training, validation, and test set.

\noindent \textbf{Multi-Source Knowledges.}
We employ TAGME, Faster R-CNN \cite{Ren15}, and GPT-turbo-3.5 to obtain textual entities, visual objects, and key phrases. respectively. The thresholds are set at 0.3 and 0.8 for textual and visual knowledge extraction of TAGME and Faster R-CNN.

\noindent \textbf{Baselines.}
We compare our model with the following representative approaches.
\textbf{BERT} \cite{DBLP:conf/naacl/DevlinCLT19} is a pre-trained model where [CLS] token is used as the textual representation to classify the claim.
\textbf{DeBERTa} \cite{deberta} and \textbf{Swin Transformer} \cite{swinv2} use DeBERTa and Swin Transformer to extract textual or visual features, and then use cosine similarity between claim and evidence to make predictions.
% \textbf{Swin Transformer} \cite{swinv2} uses  to encode visual content and gather it together with cosine similarity to verify the verdict of claims.
\textbf{GPT-3.5}\footnote{https://chat.openai.com/} leverage only textual information to verify the claim without fine-tuning.
\textbf{CLIP} \cite{DBLP:conf/icml/RadfordKHRGASAM21}  learns multimodal representations and fuses them through a cross-attention module to predict the label.
\textbf{ConcatNet} \cite{Mishra22} utilizes cosine similarity to fuse inner-modal features and concatenates textual and visual representations to obtain multimodal features.
\textbf{UofA-Truth} \cite{Dhankar22} divides the task into two sub-tasks, namely text entailment and image entailment, and makes predictions based on the permutation of results of two sub-tasks.
\textbf{Truthformers} \cite{VPV22} leverages the convolution layer as the fusion method to predict the verdict of the given claim.
\textbf{Pre-CoFact} \cite{Wang22} uses the co-attention layers to fuse multimodal contents.
\textbf{Logically} \cite{Gao22} uses a decision tree classifier with several multimodal features to make predictions.

\noindent \textbf{Implementation details.}
We use a Tesla V100-PCIE GPU with 32GB memory for all experiments and implement our model via the Pytorch framework. 
The number of attention heads is set to 4, and the number of GAT convolutional layers is 2. The batch size is 8. We set the learning rate as 2e-5. We employ DeBERTa \cite{deberta} and Swin Transformer \cite{swinv2} as the pre-trained language and visual models. 

\begin{table}
\caption{Overall results (\%) of MultiKE-GAT and comparison baselines on 5-way and 3-way fact verification tasks for FACTIFY.
  We report the weighted-average F1 score to evaluate the overall performance.
  $^\dag$ means the results are from \cite{Gao22}.
  For other baselines, we reproduce them in the same environment according to their open-source projects. 
  }
\centering
   \resizebox{0.95\linewidth}{!}{$
  \begin{tabular}{|c|l|c|c|}
    \hline
    \multicolumn{2}{|c|}{\multirow{2}{*}{\textbf{Model}}} &\multicolumn{2}{c|}{\textbf{w-F1 score}} \\ \cline{3-4}
    \multicolumn{2}{|c|}{} & \textbf{5-way} & \textbf{3-way}\\
    \hline
    &DeBERTa \cite{deberta} &  63.06  & 73.45\\
    Unimodal& Swin Transformer \cite{swinv2} & 60.70 & 69.82\\ 
    & GPT-3.5 & - & 60.93 \\\hline
    & ConcatNet \cite{Mishra22} &  66.64 & 75.77\\
    & UofA-Truth$^\dag$ \cite{Dhankar22}  & 74.80 & - \\
    Multimodal& Truthformers$^\dag$ \cite{VPV22} & 74.90 & - \\
    & Pre-CoFact \cite{Wang22}  & 75.74 & {81.85}\\
    & Logically$^\dag$ \cite{Gao22} & {77.00} & 81.00 \\
    & \textbf{MultiKE-GAT} (Ours) &  \textbf{79.64} & \textbf{83.68} \\
    \hline
\end{tabular}
$}
  \label{tab-1}
\end{table}

\begin{table}
\caption{Overall results (\%) of MultiKE-GAT and comparison baselines on MOCHEG.
 Following \cite{DBLP:conf/sigir/YaoS0CH23}, we report the F1 score of each method. 
 $^\dag$ means the results are from \cite{DBLP:conf/sigir/YaoS0CH23}.
 For other baselines, we reproduce them in the same environment according to their open-source projects. 
  }
\centering
  \begin{tabular}{|c|l|c|}
    \hline
    \multicolumn{2}{|c|}{\textbf{Model}} & \textbf{F1 score}\\
    \hline
    & BERT$^\dag$ \cite{DBLP:conf/naacl/DevlinCLT19} &  33.98\\
    Unimodal& CLIP-Text$^\dag$ \cite{DBLP:conf/icml/RadfordKHRGASAM21} & 45.18\\ 
    & CLIP-Image$^\dag$ \cite{DBLP:conf/icml/RadfordKHRGASAM21} & 40.93 \\\hline
    & CLIP$^\dag$ \cite{DBLP:conf/icml/RadfordKHRGASAM21} &  49.43 \\
    % \cite{Wang22}
    Multimodal & ConcatNet \cite{Mishra22} & 50.12  \\
    & Pre-CoFact \cite{Wang22}  & 68.45 \\
    & MultiKE-GAT (Ours) &  \textbf{70.14} \\
    \hline
\end{tabular}
  \label{tab-2}
\end{table}

\begin{table}[t]
\caption{Results (\%) of ablation study on 5-way task on FACTIFY. Multi-Knowledge denotes extracted key information and entities. Graph Fusion denotes the fusion method based on GAT. Global denotes the concatenation of global representations.}
\centering
   \resizebox{0.95\linewidth}{!}{
  \begin{tabular}{|l|c|c|}
    \hline
    \multicolumn{1}{|c|}{\textbf{Model}}& \textbf{w-F1}  & \textbf{Acc}\\
    \hline
    MultiKE-GAT &  \textbf{79.64} & \textbf{79.55}\\
    \ \ \ \  - w/o Multi-Knowlege & 73.29  ($\downarrow$ 6.35)&  73.35 ($\downarrow$ 6.20)\\
    \ \ \ \  - w/o Graph Fusion & 75.87 ($\downarrow$ 3.77) & 75.89 ($\downarrow$ 3.66)\\
    \ \ \ \  - w/o Global & 77.48 ($\downarrow$ 2.16) & 77.52 ($\downarrow$ 2.03) \\
    \hline
\end{tabular}
 }
  \label{tab-3}
\end{table}

\subsection{Results}
We performed the proposed model together with baselines on both 5-way and 3-way multimodal fact verification tasks to evaluate our model's performance on coarse-grained and fine-grained classification tasks. The results are shown in table~\ref{tab-1} and \ref{tab-2}. 
MultiKE-GAT achieves state-of-the-art performance in the multimodal fact verification tasks on FACTIFY and MOCHEG, showing the superiority of the proposed model. From the results, we have the following observations:
1) Unimodal methods have lower performance, demonstrating the necessity of incorporating multimodal information to make predictions. 
2) Large language models such as GPT-3.5 with world knowledge fail to outstandingly verify the truthfulness of claims when it is used as the classifier directly. 
3) Among multimodal approaches, MultiKE-GAT achieves better performance consistently, which indicates that infusing multi-source knowledge with graph-based fusion can learn comprehensive multimodal clues for verification.

\begin{table}[t]
\caption{Results (\%) against fusing different sources of knowledge. KP denotes key information extracted by LLMs. Ent (text/image) denotes the textual/visual entities.}
\centering
   \resizebox{0.95\linewidth}{!}{$
  \begin{tabular}{|c|l|c|c|}
    \hline
    \multicolumn{2}{|c|}{\textbf{Model}}& \textbf{w-F1}  & \textbf{Acc}\\
    \hline
    &MultiKE-GAT & 79.64 & 79.55 \\
    Multiple source&\ \ \ - w (KP, Ent (text)) & 76.74 & 76.61 \\
    &\ \ \ - w (KP, Ent (image)) & 76.87 & 76.89 \\
    &\ \ \ - w (Ent(text), Ent (image)) & 78.01 & 78.09 \\ \hline
    &\ \ \ - w KP & 74.29  &  74.35\\
    Single source&\ \ \ - w Ent (text) & 75.87 & 75.89 \\
    &\ \ \ - w Ent (image) & 74.77 & 74.81 \\    
    % \hline
    \hline
\end{tabular}
$}
  \label{tab-4}
\end{table}

\begin{table}[t]
 \caption{Results (\%) of comparison of different multimodal fusion methods on the 5-way task on FACTIFY.}
\centering
   % \resizebox{0.98\linewidth}{!}{$
  \begin{tabular}{|l|c|c|}
    \hline
    \multicolumn{1}{|c|}{\textbf{Fusion Module}}& \textbf{w-F1}  & \textbf{Acc}\\
    \hline
    Concat Fusion & 75.84 & 75.89 \\
    Self-att Fusion & 76.98 & 77.01 \\ \hline
    GCN & 77.33 & 77.34 \\
    Independent GAT & 79.31 & 79.26 \\
    \textbf{KGF} (Ours) & \textbf{79.64} & \textbf{79.55} \\
    \hline
\end{tabular}
  \label{tab-5}
\end{table}

\begin{figure*}[t]
\centering
    \subfigure[Ground-truth Label: Support Multimodal.]{
    \begin{minipage}[t]{0.43\linewidth}
    \centering
      \includegraphics[width=\linewidth]{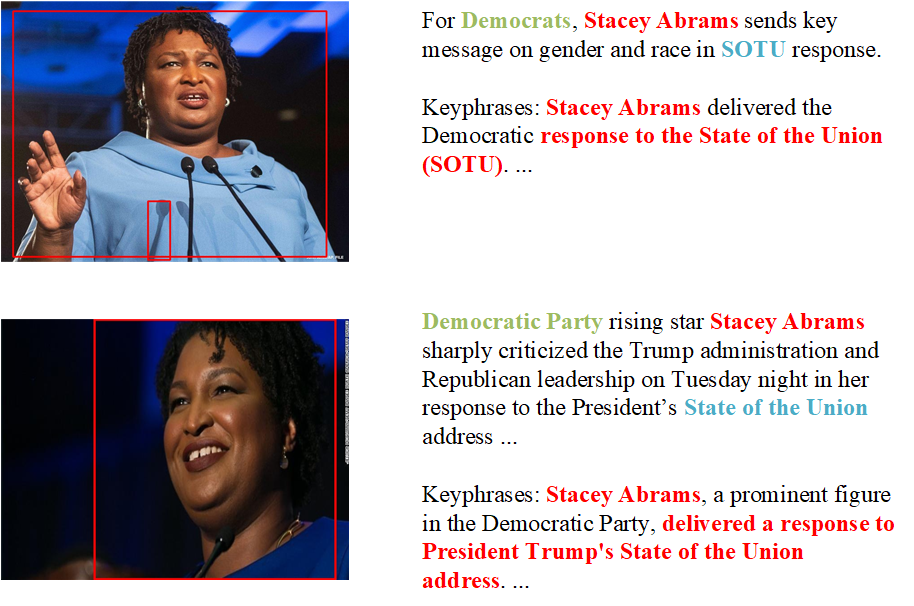}
      \label{a}
    \end{minipage}}
  \subfigure[Ground-truth Label: Support Multimodal]{
  \begin{minipage}[t]{0.43\linewidth}
      \centering
      \includegraphics[width=\linewidth]{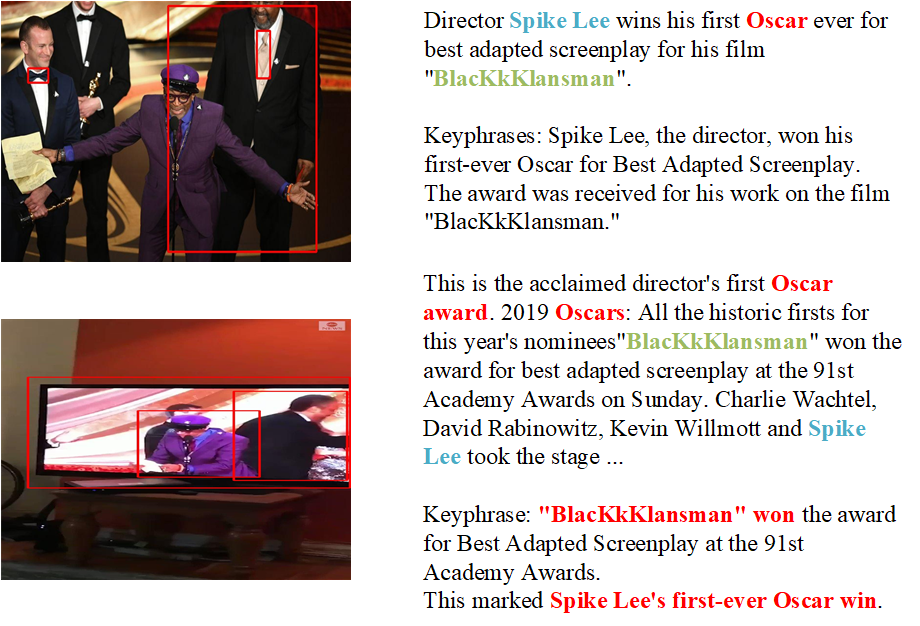}
      \label{b}
    \end{minipage}}
  \caption{Two cases of FACTIFY. Red rectangles demonstrate objects and bold words are entities and key information.}
  \label{f:4}
\end{figure*}

\subsection{Ablation study}
Table \ref{tab-3} shows the results of MultiKE-GAT and its variants on FACTIFY. 
When removing multi-source knowledge, the results of  w/o Multi-Knowledge decline significantly. This proves the effectiveness of multi-source knowledge.
Besides, we obtain inferior performance when removing the knowledge-oriented fusion module and simply fusing multimodal knowledge with concatenation.  
Furthermore, the performance of w/o Global in the KGF module also drops, indicating the important role of global nodes in multimodal fusion.

\subsection{The Effect of multi-source knowledge}

{Table \ref{tab-4} shows comparison results of MultiKE-GAT using different sources of knowledge such as textual entities, visual objects, and key phrases. MultiKE-GAT obtains the best performance when all of these three kinds of knowledge are utilized. Multi-source knowledge-based methods achieve better performance than single-source knowledge-based methods. It demonstrates that multi-source knowledge can leverage diverse information types and sources to validate claims across different modalities.}

\subsection{Comparison of different fusion modules}

{In this part, we further compare the proposed KGF module with several representative multimodal fusion modules including concatenation fusion, self-attention fusion, and three graph-based fusion methods. 
The results on FACTIFY are shown in table \ref{tab-5}. 
Our KGF module achieves better performance in terms of w-F1 and accuracy scores. 
1) Compared with graph-based fusion modules, Concat and Self-att fusion modules usually focus on coarse-grained patterns and learn shallow interactions between modalities, leading to relatively inferior performance.
2) Different from other graph-based fusion modules, our KGF learns complex interactions and effectively integrates fine-grained knowledge clues across different modalities with the guidance of global nodes.  
It adeptly addresses inconsistencies and filters out noise present within diverse knowledge sources.}

\subsection{Case study}
Figure \ref{f:4} shows some cases from the test set of FACTIFY. The two samples belong to the Support Multimodal category and are correctly classified by our model but misjudged by Pre-CoFact. 
{1) The images of claim and evidence in figure \ref{a} depict the same person despite the different clothes, expressions, and degrees of brightness. However, due to the ignorance of fine-grained visual information, the Pre-CoFact model wrongly classifies it into \textit{Support Text}.
2) For Figure \ref{b}, the evidence image is shot on television and is different from the claim image. Thus
it's challenging to verify the relation between two images using only coarse-grained features. Pre-CoFact incorrectly classifies it as \textit{Insufficient Multimodal}.
However, our model focuses on fine-grained clues. It not only easily verifies the claim text's authenticity based on nuanced details such as matching names, locations, and events, but also identifies that the two images are highly similar, containing the same individuals and events.
3) Each keyphrase extracted by LLM implicitly divides claims and evidence into multiple individual factual points. This multi-source knowledge facilitates the model to focus on effective fine-grained information to help make predictions. }

\section{Conclusion}
\label{sec:conc}

In this work, we propose MultiKE-GAT to extract fine-grained entity features and perform comprehensive multimodal fusion for fact verification.
Besides, we design a novel Knowledge-aware Graph Fusion architecture to eliminate the inconsistency and noise introduced by irrelevant entities. 
The experimental results on FACTIFY show that MultiKE-GAT is capable of effectively dealing with multimodal fact verification tasks in comparison with other competitive methods. These results highlight the effectiveness and superiority of our proposed model.
For future work, we will focus on exploring how external knowledge can be used to provide better explanations to further enhance multimodal fact verification.

\section*{Acknowledgment}

This work was supported by the National Key Research and Development Program of China (No. 2022YFC3302102), and the Postdoctoral Fellowship Program of China Postdoctoral Science Foundation (No. GZC20232969).

\bibliographystyle{IEEEtran}
\bibliography{icme2023template}

\begin{thebibliography}{10}
\providecommand{\url}[1]{#1}
\csname url@samestyle\endcsname
\providecommand{\newblock}{\relax}
\providecommand{\bibinfo}[2]{#2}
\providecommand{\BIBentrySTDinterwordspacing}{\spaceskip=0pt\relax}
\providecommand{\BIBentryALTinterwordstretchfactor}{4}
\providecommand{\BIBentryALTinterwordspacing}{\spaceskip=\fontdimen2\font plus
\BIBentryALTinterwordstretchfactor\fontdimen3\font minus \fontdimen4\font\relax}
\providecommand{\BIBforeignlanguage}[2]{{%
\expandafter\ifx\csname l@#1\endcsname\relax
\typeout{** WARNING: IEEEtran.bst: No hyphenation pattern has been}%
\typeout{** loaded for the language `#1'. Using the pattern for}%
\typeout{** the default language instead.}%
\else
\language=\csname l@#1\endcsname
\fi
#2}}
\providecommand{\BIBdecl}{\relax}
\BIBdecl

\bibitem{GuoSV22}
\BIBentryALTinterwordspacing
Z.~Guo, M.~S. Schlichtkrull, and A.~Vlachos, ``A survey on automated fact-checking,'' \emph{{TACL}}, vol.~10, pp. 178--206, 2022. [Online]. Available: \url{https://doi.org/10.1162/tacl\_a\_00454}
\BIBentrySTDinterwordspacing

\bibitem{Roy21}
A.~Roy and A.~Ekbal, ``Mulcob-mulfav: Multimodal content based multilingual fact verification,'' in \emph{{IJCNN}}, 2021, pp. 1--8.

\bibitem{Singhal22}
\BIBentryALTinterwordspacing
S.~Singhal, T.~Pandey, S.~Mrig \emph{et~al.}, ``Leveraging intra and inter modality relationship for multimodal fake news detection,'' in \emph{{WWW}}, 2022, pp. 726--–734. [Online]. Available: \url{https://doi.org/10.1145/3487553.3524650}
\BIBentrySTDinterwordspacing

\bibitem{Wu21}
\BIBentryALTinterwordspacing
Y.~Wu, P.~Zhan, Y.~Zhang, L.~Wang, and Z.~Xu, ``Multimodal fusion with co-attention networks for fake news detection,'' in \emph{Findings of {ACL}}, Aug. 2021, pp. 2560--2569. [Online]. Available: \url{https://aclanthology.org/2021.findings-acl.226}
\BIBentrySTDinterwordspacing

\bibitem{Qian21}
S.~Qian, J.~Wang, J.~Hu, Q.~Fang, and C.~Xu, ``Hierarchical multi-modal contextual attention network for fake news detection,'' in \emph{SIGIR}, 2021, pp. 153–--162.

\bibitem{Mishra22}
\BIBentryALTinterwordspacing
S.~Mishra, S.~S, A.~Bhaskar \emph{et~al.}, ``{FACTIFY:} {A} multi-modal fact verification dataset,'' in \emph{Workshop on DE-FACTIFY{@}{AAAI}}, vol. 3199, 2022. [Online]. Available: \url{https://ceur-ws.org/Vol-3199/paper18.pdf}
\BIBentrySTDinterwordspacing

\bibitem{Gao22}
\BIBentryALTinterwordspacing
J.~Gao, H.~Hoffmann, S.~Oikonomou \emph{et~al.}, ``Logically at factify 2022: Multimodal fact verfication,'' in \emph{Workshop on DE-FACTIFY{@}{AAAI}}, vol. 3199, 2022. [Online]. Available: \url{https://ceur-ws.org/Vol-3199/paper6.pdf}
\BIBentrySTDinterwordspacing

\bibitem{Dhankar22}
\BIBentryALTinterwordspacing
A.~Dhankar, O.~Za{\"{\i}}ane, and F.~Bolduc, ``Uofa-truth at factify 2022 : {A} simple approach to multi-modal fact-checking,'' in \emph{Workshop on DE-FACTIFY{@}{AAAI}}, vol. 3199, 2022. [Online]. Available: \url{https://ceur-ws.org/Vol-3199/paper10.pdf}
\BIBentrySTDinterwordspacing

\bibitem{Wang22}
W.~Wang and W.~Peng, ``Team yao at factify 2022: Utilizing pre-trained models and co-attention networks for multi-modal fact verification,'' in \emph{Workshop on DE-FACTIFY{@}{AAAI}}, vol. 3199, 2022.

\bibitem{Singhal20}
S.~Singhal, A.~Kabra, M.~Sharma, R.~R. Shah, T.~Chakraborty, and P.~Kumaraguru, ``Spotfake+: A multimodal framework for fake news detection via transfer learning,'' \emph{AAAI}, vol.~34, no.~10, pp. 13\,915--13\,916, 2020.

\bibitem{QiCao21}
\BIBentryALTinterwordspacing
P.~Qi, J.~Cao, X.~Li \emph{et~al.}, ``Improving fake news detection by using an entity-enhanced framework to fuse diverse multimodal clues,'' in \emph{{MM}}, 2021, pp. 1212--1220. [Online]. Available: \url{https://doi.org/10.1145/3474085.3481548}
\BIBentrySTDinterwordspacing

\bibitem{tan2023evidencebased}
X.~Tan, B.~Zou, and A.~T. Aw, ``Evidence-based interpretable open-domain fact-checking with large language models,'' 2023.

\bibitem{DBLP:conf/sigir/YaoS0CH23}
B.~M. Yao, A.~Shah, L.~Sun, J.~Cho, and L.~Huang, ``End-to-end multimodal fact-checking and explanation generation: {A} challenging dataset and models,'' in \emph{{SIGIR}}, 2023, pp. 2733--2743.

\bibitem{Ren15}
S.~Ren, K.~He, R.~Girshick, and J.~Sun, ``Faster r-cnn: Towards real-time object detection with region proposal networks,'' in \emph{NeurIPs}, 2015.

\bibitem{PengQ19}
\BIBentryALTinterwordspacing
Y.~Peng and J.~Qi, ``Cm-gans: Cross-modal generative adversarial networks for common representation learning,'' \emph{{ACM} Trans. Multim. Comput. Commun. Appl.}, vol.~15, no.~1, pp. 22:1--22:24, 2019. [Online]. Available: \url{https://doi.org/10.1145/3284750}
\BIBentrySTDinterwordspacing

\bibitem{DBLP:conf/naacl/DevlinCLT19}
J.~Devlin, M.~Chang, K.~Lee, and K.~Toutanova, ``{BERT:} pre-training of deep bidirectional transformers for language understanding,'' in \emph{{NAACL-HLT}}, 2019, pp. 4171--4186.

\bibitem{deberta}
\BIBentryALTinterwordspacing
P.~He, X.~Liu, J.~Gao, and W.~Chen, ``Deberta: Decoding-enhanced bert with disentangled attention,'' in \emph{ICLR}, 2021. [Online]. Available: \url{https://openreview.net/forum?id=XPZIaotutsD}
\BIBentrySTDinterwordspacing

\bibitem{swinv2}
\BIBentryALTinterwordspacing
Z.~Liu, H.~Hu, Y.~Lin, Z.~Yao, Z.~Xie, Y.~Wei, J.~Ning, Y.~Cao, Z.~Zhang, L.~Dong, F.~Wei, and B.~Guo, ``Swin transformer {V2:} scaling up capacity and resolution,'' \emph{CoRR}, vol. abs/2111.09883, 2021. [Online]. Available: \url{https://arxiv.org/abs/2111.09883}
\BIBentrySTDinterwordspacing

\bibitem{DBLP:conf/icml/RadfordKHRGASAM21}
A.~Radford, J.~W. Kim, C.~Hallacy, A.~Ramesh, G.~Goh, S.~Agarwal, G.~Sastry, A.~Askell, P.~Mishkin, J.~Clark, G.~Krueger, and I.~Sutskever, ``Learning transferable visual models from natural language supervision,'' in \emph{{ICML}}, 2021, pp. 8748--8763.

\bibitem{VPV22}
C.~B. S.~N. V, P.~Potluri, and R.~Vijjali, ``Truthformers at factify 2022 : Evidence aware transformer based model for multimodal fact checking,'' in \emph{Workshop on DE-FACTIFY{@}{AAAI}}, vol. 3199, 2022.

\end{thebibliography}

\end{document}